# An unprecedented synergy of high-temperature tensile strength and ductility in a NiCoCrAlTi high-entropy alloy

Hongmin Zhang[a], Fanchao Meng[a*], Haoyan Meng[a], Yang Tong[a], Peter K. Liaw[b], Xiao Yang[c], Lei Zhao[d], Haizhou Wang[d], Yanfei Gao[b*], Shuying Chen[a*]

[a] *Institute for Advanced Studies in Precision Materials, Yantai University, Yantai, Shandong 264005, China*

[b] *Department of Materials Science and Engineering, The University of Tennessee, Knoxville, TN 37996, USA*

[c] *Key Laboratory of Cryogenics, Technical Institute of Physics and Chemistry, Chinese Academy of Sciences, Beijing 100190, China*

[d] *Beijing Advanced Innovation Center for Materials Genome Engineering, Beijing Key Laboratory of Metal Materials Characterization, Central Iron and Steel Research Institute, Beijing 100081, China*

## Abstract

The present work reported a novel $L1_2$-strengthening NiCoCrAlTi high entropy alloy (HEA) with an outstanding synergy of tensile strength and ductility at both ambient and high temperatures. Transmission electron microscopy (TEM) characterization revealed a high density of rod-like and spheroidal $L1_2$ precipitates distributing in the micro/nanograins and non-recrystallized regions in the annealed specimens. The tremendously high yield stress, ultimate tensile stress (UTS), and ductility of the HEA at 600 °C were ~1060 MPa, 1271 MPa, and 25%, respectively, which were significantly superior to most reported HEAs and Co- and Ni-based superalloys to date. Systematic TEM analysis unveiled that the cooperation among $L1_2$ precipitation, extensive stacking faults (SFs), deformation twins (DTs), immobile Lomer-Cottrell (L-C) locks

---
* Corresponding authors.
  E-mail addresses: sychen@ytu.edu.cn (S.Y. Chen), mengfanchao@ytu.edu.cn (F.C. Meng), ygao7@utk.edu (Y.F. Gao)

formed from interactions between SFs and SFs/DTs, hierarchical SFs/DTs networks, as well as hetero-deformation-induced strengthening dominated the plastic deformation at 600 °C. Such a unique deformation mechanism enabled extremely high tensile strength and sustained ductility of the HEA at a high temperature.

## 1. Introduction

A new type of multi-principal component HEA possesses inherent global disorder, high lattice distortion, composition and structural complexity, and a "cocktail" effect, which is quite different from the traditional alloy design paradigm [1]. The emergence of HEAs gives a huge degree of freedom to regulate the structure and properties of the alloy, which provides a new design strategy and an extremely broad imagination space for breaking through the synergy of strength and ductility [2]. In recent years, HEAs have shown many advantages in high-temperature environments [3, 4]. Among the numerous types of HEAs, there are three main types of HEAs with broad high-temperature application prospects: refractory high entropy alloys [5, 6], eutectic high entropy alloys [7-10], and precipitation-strengthened high entropy alloys (PSHEAs) [3, 10].

PSHEAs are one of the most promising candidates for elevated-temperature applications [10]. The formation of massive precipitation generated strong obstacles to hinder the dislocation motion, rendering them excellent resistance against softening and grain growth at high temperatures [11]. Additionally, alloying of multiple principal elements with different atomic size and physical properties significantly increases activation energies for elemental diffusion, which contribute to enhanced resistance to coarsening for the nano-precipitate when compared with many other conventional Ni-based superalloys [12, 13]. Lu et al. studied the mechanical properties of interstitial HEA (iHEA) of $Fe_{19.84}Mn_{19.84}Co_{19.84}Cr_{19.84}Ni_{19.84}C_{0.8}$. The good synergy of the yield stress of 670 MPa and ductility of 16% was obtained at 600 °C, which is attributed to the interfacial nanophase formation of 9R phase and elongated nano-carbides, stabilizing the nanotwin in the non-recrystallized regions [14]. In order to avoid the GB embrittlement in $Ni_{30}Co_{13}Fe_{15}Cr_6Al_6Ti_{0.1}B$ alloy during a high-temperature tensile test, a duplex aging process can be adopted to suppress the formation of undesired GB heterostructures of cellular $L1_2$ precipitates and Heusler phase, leading to a twofold enhancement of tensile ductility [12].

Even though the SFs and twinning-induced plasticity (TWIP) usually served as the deformation mechanism at ambient and cryogenic temperatures, their strengthening effects had been rarely observed at high temperatures, whereas dynamic recovery is commonly seen. It has been shown in both traditional and high entropy alloys that twinning strength is rather athermal. Therefore, as the common observations, the high strength at lower temperatures is often accompanied by lots of twinning activities, but the low strength at higher temperatures usually does not reach the twin strength. Recent studies observed the obvious formation of SFs and revealed their crucial strengthening effects in PSHEAs when deformed at elevated temperatures. For instance, excellent high-temperature performance can be achieved in the $(Ni_2Co_2FeCr)_{92}Al_4Nb_4$ HEAs developed by Zhao et al. [15], especially at temperatures above 600 °C. The strength of the $Al_4Nb_4$ HEA is stronger than that of most previously studied solid-solution-strengthened HEA, which can be attributed to the shearing of $L1_2$ precipitates by stacking faults as the dominant deformation mode. The $Ni_{39.9}Co_{20}Fe_{10}Cr_{20}Al_6Ti_4B_{0.1}$ HEA designed by Hou et al. [16] exhibits an excellent strain hardening capability with an outstanding strength-ductility combination at 600 °C, which can be ascribed to the dominant deformation mechanism of SFs formed in the matrix with intermittent configurations. Zhang et al. [17] attributed the superior tensile strengths of Fe-28.2Ni-17Co-11Al-2.5Ta-0.04B HEA at 600 and 800°C to the $L1_2$ precipitation strengthening and the dislocation dissociation and the consequent formation of SFs. Previous studies suggested that the SFs formation could be one of the dominant deformation mechanisms of high-temperature tensile loading. However, the improved strength and ductility in the previous PSHEAs are still very limited when compared with the traditional Ni- and Co-based superalloys. Inspired by the previous reports, more strengthening mechanisms, e.g., extensive DTs and SFs, need to be introduced at high temperatures to maintain the tremendous synergy of strength and ductility, similar to the cryogenic tensile behavior. In the present work, we design a novel HEA, $Ni_{33.3}Co_{33.3}Cr_{23.4}Al_5Ti_5$ with reduced Cr, increased Ni and Co to avoid the σ formation, lower the SFE, as well as enhance the $L1_2$ nucleation. Based on such alloy design, the precipitation hardening, DTs, SFs, and L-C locks could cooperate effectively to achieve outstanding comprehensive mechanical properties at 600 °C.

## 2. Materials and methods

The alloy button ingots with a nominal composition of $Ni_{33.3}Co_{33.3}Cr_{23.4}Al_5Ti_5$ (at. %) were arc-melted with the elements Co, Cr, Ni, Al, and Ti (> 99.9 wt.%) under high purity argon atmosphere. To promote thorough mixing, each button ingot was flipped and re-melted at least six times. All button ingots were finally cast into a copper mold with a sectional dimension of $80 \times 8 \times 8$ mm$^3$. The as-cast ingots were homogenized at 1473 K for 24 h followed by water quenching to reduce the dendritic segregation. The homogenized alloys were then rolled in steps at ambient temperature along the longitudinal direction to a total thickness reduction of ~83% (~ 1.5 mm). Finally, the as-rolled sheets were annealed at 1153 K for 3 h followed by water quenching.

Flat dog-bone-shaped specimens with a gauge length of 20 mm, a width of 3.7 mm, and a thickness of ~1.4 mm were prepared by electro-discharge machining for uniaxial tensile tests. The tensile tests were all conducted and duplicated 3 times on a SENS CMT5105 tensile test machine. A fixed strain rate of $1 \times 10^{-3}$ /s was selected for the conventional tensile tests conducted at room temperature (RT) and 600 °C. The strain measurement was carried out according to the marked spacing of the gage length.

The microstructural characterization before and after the tensile test was performed by using a Tescan Amber scanning electron microscope (SEM), equipped with an Oxford electron backscattered diffraction (EBSD). TEM analysis was carried out with an FEI Tecnai F30 operating at 200 kV. Phase diagram calculation based on CALPHAD was used to calculate the fraction of the equilibrium phase. In the present work, the TCHEA4 database was adopted and calculated, as shown in Fig. 1(a).

## 3. Results and discussion

Phase stability of the CoCrNiTiAl alloy was evaluated by CALPHAD method. Fig. 1(a) shows the equilibrium phase fractions as a function of temperature. The two-phase region, consisting of equilibrium γ and γ' phases, exists in the temperature range from 880 to 1080 °C. In order to acquire the largest volume fraction of L1$_2$ to realize the maximum strengthening effect, while excluding the formation of σ phase, the annealing temperature was chosen as 880 °C. The backscattering electron (BSE) image in Fig. 1(b) presents the heterogeneous structure of sample annealing at 880 °C for 3 h, which involved in regions: recrystallized regions (RG) with fine grains and non-recrystallized region (NR) with deformed structure (as displayed in the inset). To reveal detailed microstructure in both regions, the etched specimen is displayed in Fig. 1(c). The interface of the two regions is outlined by the yellow dashed line. Totally different microstructure, flat and flowered features were observed in NR and RG, respectively. Magnified Figs. 1(d) and (e) were illustrated to closely capture the structure in both regions, as indicated by the red rectangle in Fig. 1(c). A large number of spheroidal precipitates in the nano-scale were homogeneously formed in the whole NR, which is associated with continuous changes in the matrix composition [18]. By contrast, numerous rod-like precipitates were shown in the RG. The rod-like precipitates refer to the heterogeneous nucleation of discontinuous precipitation (DP), which was promoted by the migration of grain boundaries/lattice diffusion [19].

Moreover, EBSD measurement was carried out to confirm the recrystallized state in the aged specimen, as shown in Fig. 1(f). The inverse pole figure (IPF) displayed nano-sized and micro-sized recrystallized grains in the RG, and the right side was indicated as NR. The kernel average misorientation (KAM) of the inset also demonstrated that RG and NR were with a low and high density of geometrically necessary dislocation (GND), respectively. The formation of the NR and RG microstructure in Fig. 1 suggests the contribution from the hetero-deformation-induced (HDI) strengthening [20]. Due to the strength difference, there will be accumulation of GNDs near the boundary of these two regimes.

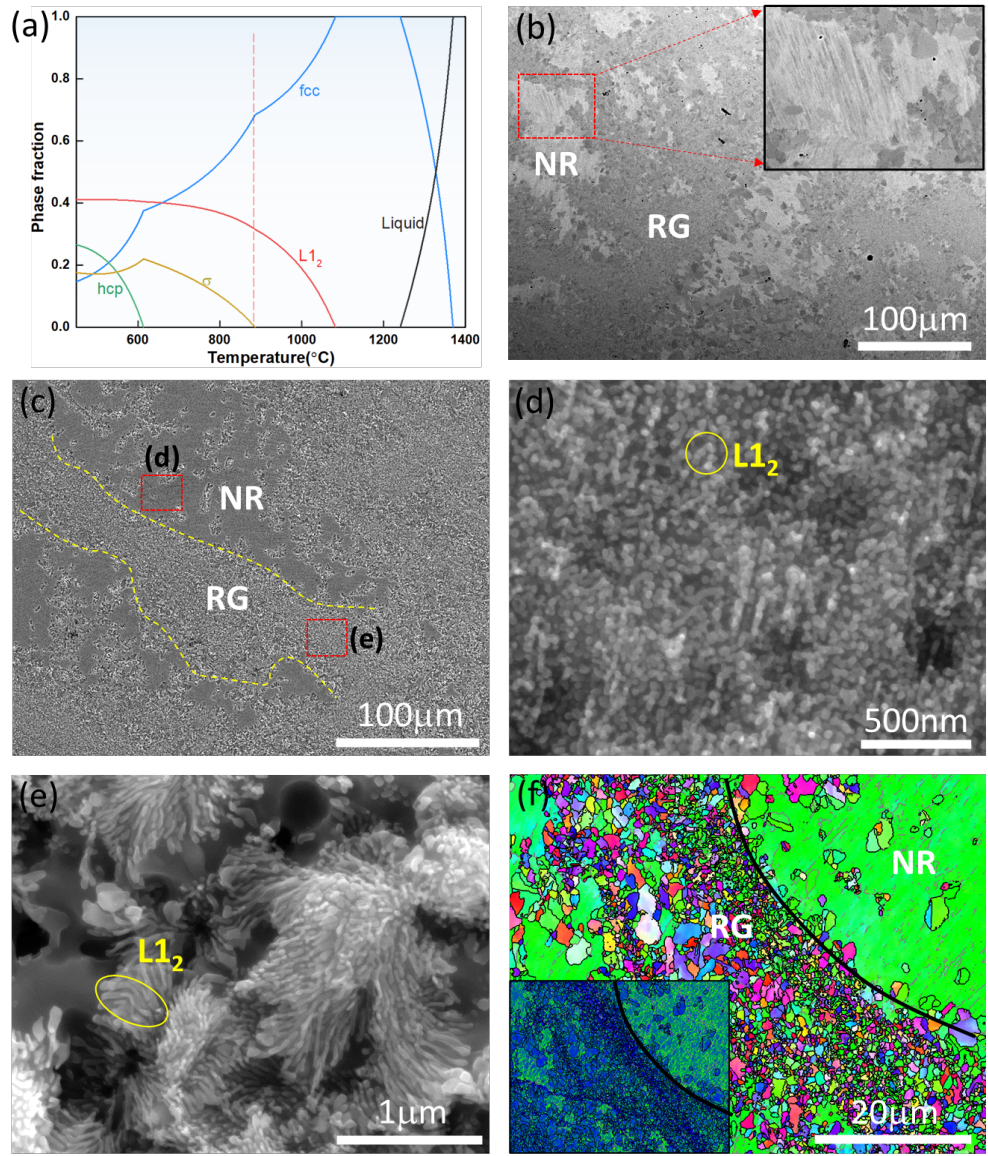

**Fig. 1.** (a-d) SEM of annealed specimen at 880 °C for 3 h. The recrystallization grain and non-recrystallization region are marked as RG and NR, respectively. The interface between recrystallized and non-recrystallized regions is marked by a yellow dashed line in the (e, f) IPF and KAM images of specimens including recrystallized and non-recrystallized regions.

Fig. 2 presents the detailed microstructural characterization of the aged NiCoCrAlTi sample. The bright-field (BF) TEM image in Fig. 2(a) shows a fine-grained microstructure after annealing at 880 °C for 3 h. Micro- and nano-size grains were distinguished by the white dashed curves. The SAED pattern in the inset of Fig. 2(a) indicates the fcc structure of the matrix. The scanning TEM (STEM) image in Fig. 2(b) displayed the spherical precipitates with an average size of 50 nm uniformly distributed in the NRs. Whilst the dark field (DF) image captured near [110] zone axis in Fig. 2(c) clearly depicts the $L1_2$ non-rod with an elongated cylindrical shape in the RG. The growth direction of these rods (the long axis of the cylinder) does not lie along any specific crystallographic direction but instead tries to be locally normal to the diffusion front sweeping the grain, which has been reported in detail previously [18, 19]. SAED pattern of both precipitates in the inset with additional faint points confirms a superlattice $L1_2$ structure. A close view of both precipitates through the high-resolution TEM (HRTEM) along [110] zone axis was displayed in Figs. 2(d) and (g), respectively. The Fast Fourier transformation (FFT) images of the matrix and precipitates revealed the fcc and $L1_2$ structure as well. Moreover, the HRTEM coupled with the corresponding FFT images displayed the coherent interface between the matrix and the precipitates. The lattice constant of fcc matrix, rod-like, and spherical $L1_2$ precipitates was calculated as 0.355, 0.351, and 0.352 nm, respectively, according to the TEM characterization. Furthermore, the lattice mismatch between fcc matrix and the two types of precipitates was determined as 1.12% and 0.8%, respectively, suggesting that both types of precipitates are highly coherent with the fcc matrix. Figs. 2(j-o) shows the example of the composition STEM-EDS mapping of spherical precipitates in the NR. The precipitates in the NR region are enriched with Ni, Al, and Ti while keeping a relatively low concentration of Co and Cr. Ni, Al, and Ti have a very strong tendency to partition into the $L1_2$ nanoparticles, whilst Co is partially depleted and Cr is largely depleted from the nanoparticles.

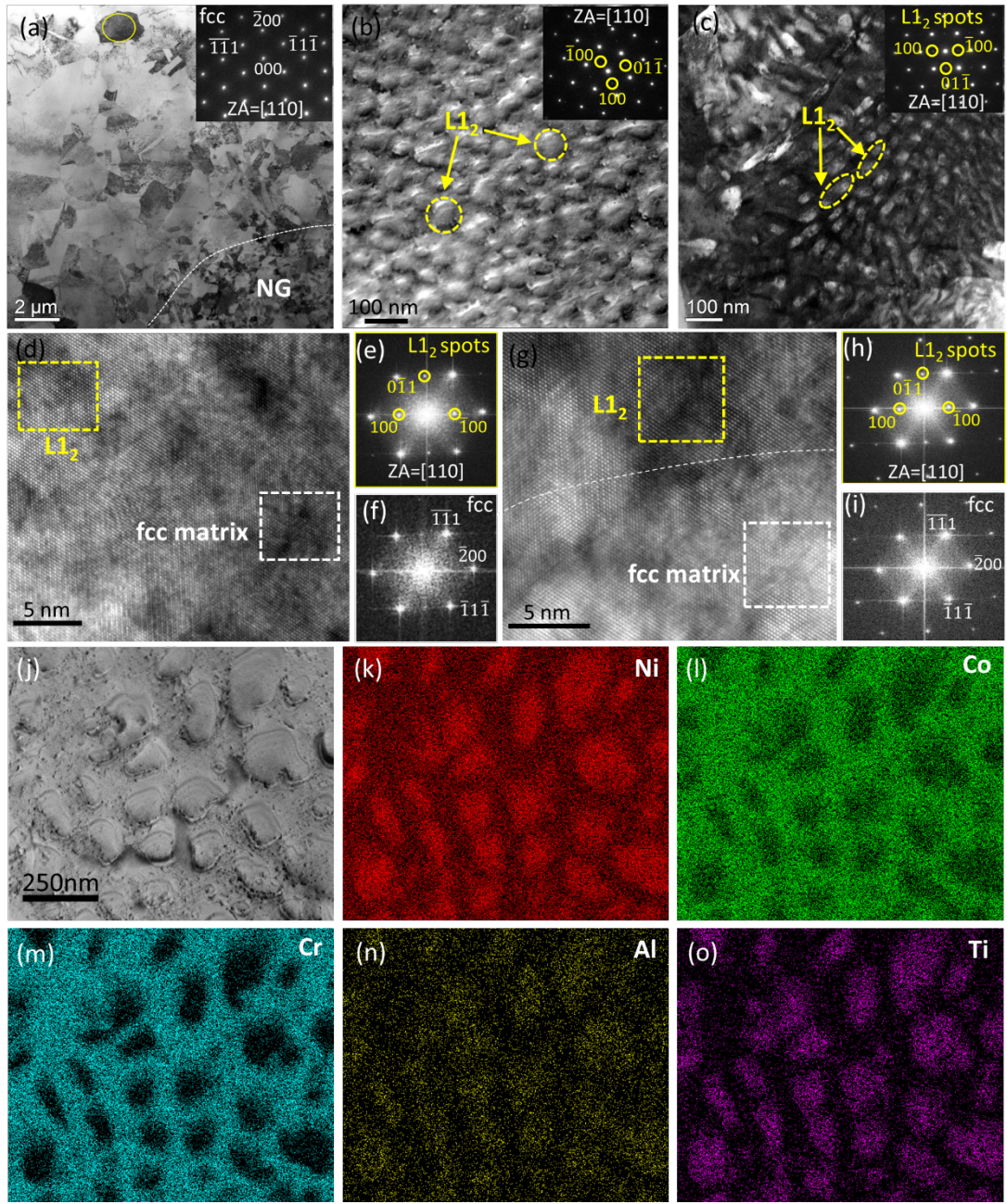

**Fig. 2.** TEM image of the aged specimen. (a) BF image of a large area, (b) STEM image of spherical precipitates taken from NR, (c) DF image of rod-like precipitates taken from RG, (d) HRTEM image taken from (b), (e, f) FFT taken from yellow and white rectangle from (d), (g) HRTEM image taken from (c), (h, i) FFT taken from yellow and white rectangle from (g). (j-o) Example of TEM-EDS image taken from round precipitates in NR.

To examine the temperature dependence of the mechanical properties, quasi-static tensile loading (strain rate $\sim 10^{-3}$/s) was conducted for the NiCoCrAlTi alloys at both ambient and elevated temperatures. The engineering stress-strain plots of the conditions (Fig. 3(a)) show that the alloy deformed at RT exhibits a yield strength ($\sigma_y$) of ~1300 MPa, UTS($\sigma_{UTS}$) of ~1610 MPa, and 14% plastic strain ($\varepsilon$) to failure. In addition to the excellent RT mechanical properties, a rather superior combination of strength and ductility with a $\sigma_y$ of ~1060 MPa, $\sigma_{UTS}$ of ~1280 MPa, and decent tensile ductility of 25% are obtained at temperatures as high as 600 °C, which is extremely superior to that of high-temperature engineering-based superalloys and other HEAs. Three duplicate tensile experiments were carried out at 600°C and similar results were obtained (see Fig. S1 in the Supplementary Information for details). Fig. 3(b) presents the work-hardening behavior for the two temperatures. The sample deformed at 600 °C exhibits a two-stage characteristic, while the specimen deformed at RT illustrates an abrupt increase in stage II, followed by a sudden drop. A $\sigma_{UTS}$ of over 1.6 GPa was achieved in the MEA when tested at RT, which can be attributed to the exceptional strain-hardening capabilities at RT. Fig. 3(c) presents the comparison of the product of $\sigma_{UTS}*\varepsilon$ versus yield strength at 600 °C. The studied alloy displayed an outstanding combination of strength and ductility at 600 °C, when compared with other high-performance alloys, including conventional $L1_2$-strengthened Ni-based superalloys [21-25], Co-based superalloys [26, 27], and other HEAs [7, 8, 14-17, 28-33]. Specifically, the yield strengths of the NiCoCrTiAl HEA are, respectively, ~ 1.7 and 4.2 times higher than those of well-known advanced Inconel 625 and equiatomic CrMnFeCoNi Cantor HEA at 600 °C. Such significantly enhanced mechanical properties are related to the unique deformation mechanism in the present HEA. Our designed alloys demonstrate a superior strength-ductility combination in a wide range of temperatures, from ambient to elevate temperatures.

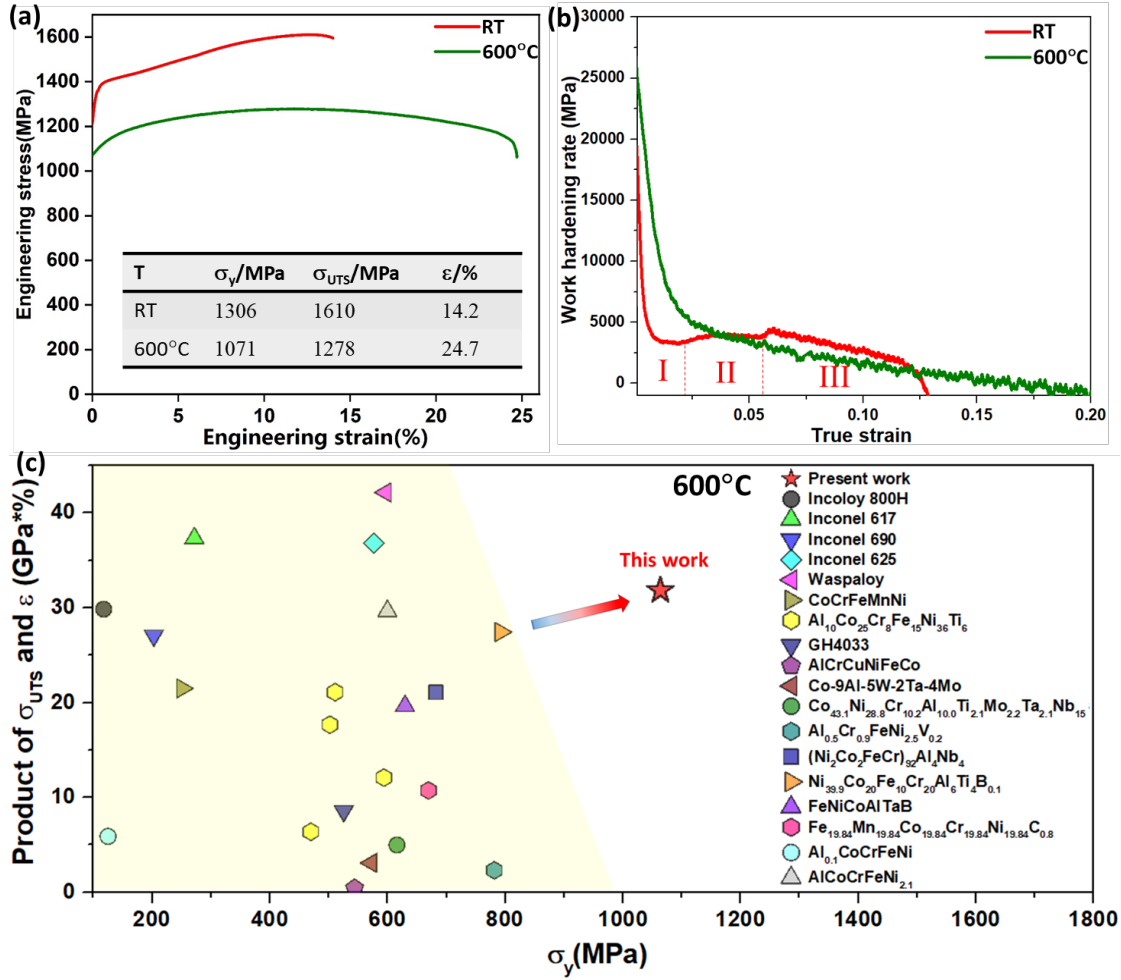

**Fig. 3.** (a) tensile engineering stress-strain tested at RT and 600 °C, (b) the corresponding work hardening rate vs. true strain, (c) the comparison of the product of $\sigma_{UTS}*\varepsilon$ vs. $\sigma_y$ [7, 8, 14-17, 21-33].

The tensile behavior of the NiCoCrAlTi HEA is fascinating at both RT and elevated temperatures of 600 °C, maintaining an outstanding synergy of high strength and ductility (Fig. 3). To further discern the immanent deformation mechanisms underlying the unusual strain hardening and tensile ductility of NiCoCrAlTi alloy at RT and 600 °C, postmortem TEM characterization was carried out in samples subjected to deformation at RT and 600 °C in Figs. 4 and 5, respectively. Figs. 4(a) and (b) presented the numerous deformation twins and SFs, which were confirmed by the SAED pattern in Fig. 4(c). The L1$_2$ phase was also detected by the yellow dash circle. The HRTEM images in Figs. 4(d-f) reveal the extensive presence and interaction of stacking faults and deformation twins. The extensive intersection of two SFs illustrated by the inserted FFT image with crossing diffraction fringes in Fig. 4(d-f) generates an immobile L-C locks structure. The production of stable L-C locks is developed by the junction reactions of a/6<112> partial dislocations moving on intersecting {111} planes, which is also observed in some other fcc alloys deformed at a cryogenic temperature [34, 35]. Additionally, most of these DTs are composed of numerous SFs, as indicated by the yellow arrow in Figs. 4(a) and (b). The HRTEM micrograph in Fig. 4(e) shows an example of twin boundaries decorated with abundant SFs, as revealed by the streaking lines in the inserted FFT pattern.

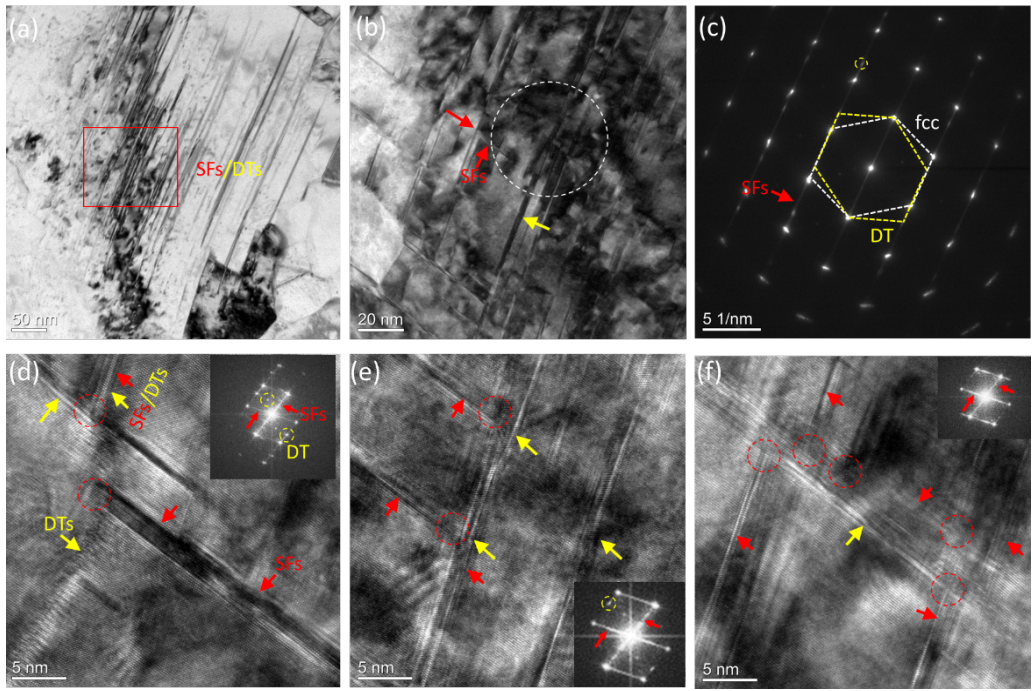

**Fig. 4**. TEM characterization of specimen deformed at RT (a) BF image revealing the DTs and SFs; (b) magnified DTs and SFs; (c) SAED pattern of matrix and DT; (d-f) HRTEM images reveal the presence of stacking faults and twins. The FFT image with crossing diffraction fringes (red arrow) in the inset of (d-f) reveals the presence of intersected SFs.

Two locations of failed specimens deformed at 600°C were chosen to systematically examine the deformation microstructure in detail, as displayed in Figs.

5(a-d) and Figs. 5(e-h). Fig. 5(a) shows a high density of tangled dislocations and a low density of slip bands, which were determined as DTs by the SAED pattern in Fig. 5(b). The corresponding DF image taken from the weak spot indicated by the red circle also confirmed the appearance of DTs with bright contrast (Fig. 5(c)). Although most SFs are parallel along one direction, i.e., along twin boundaries, Fig. 5(a) captures another set of high-density SFs on $(1\bar{1}1)$ planes forming an intersection angle of 72° relative to the SFs and DTs in the other direction. Similar to the deformation at RT, numerous intersections of SFs and DTs/SFs lead to the formation of hierarchical nano-spaced SF networks and immobile L-C locks [34, 35].

In the second area of the deformed sample, the typical planar glide of numerous SFs is activated, forming the arrays of planar SFs in two non-coplanar slip systems, as seen in Fig. 5(e). A closer observation in Fig. 5(f) displayed that most parallel SFs were characterized along $(11\bar{1})$ slip planes. Another set of SFs along $(1\bar{1}1)$ slip plane was also observed by the red arrows. The enlarged view in Fig. 5(g) closely captured the abundant intersected SFs nano-stripes on two {111} slip systems. The nature of SFs is further determined by HRTEM in Fig. 5(h), which exhibits an example of two sets of SFs interacting with each other and forming the so-called "SF parallelepipeds" structure, as indicated by the white dash line. The emergence of substantial SFs implies that the NiCoCrAlTi alloy might possess an intrinsically low SF energy. Thus, these SFs are related to the formation and movement of Shockley partial dislocations in the alloys with low SFE. The interior angle between the two sets of intersected stacking faults is around 72°, which is consistent with that of {111} close-packed planes in the fcc-type unit cell [33, 36]. Moreover, no hcp or σ phase formed during the tensile test at 600℃, which may be due to the short time before the sample failed.

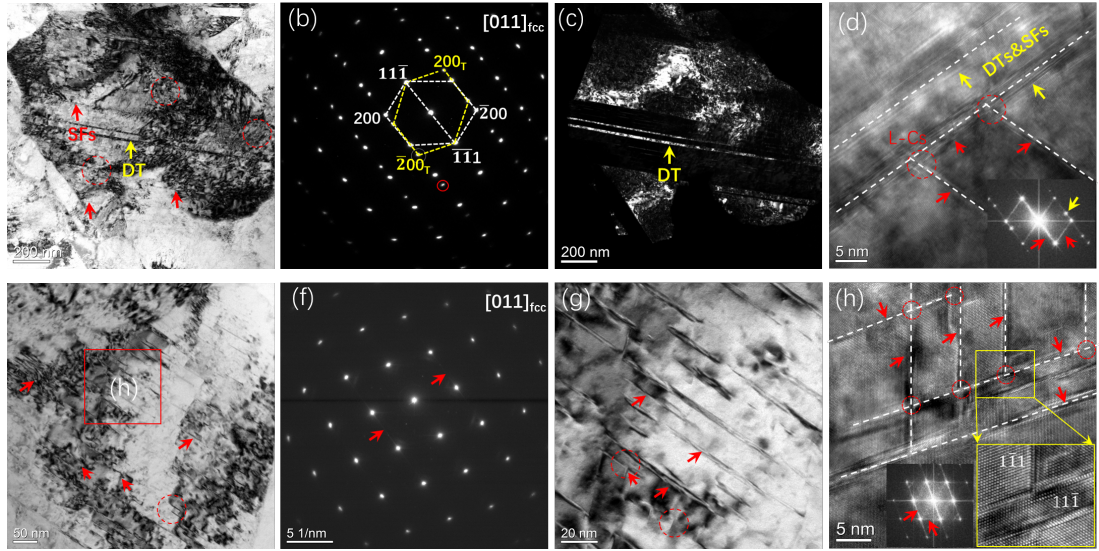

**Fig. 5.** The TEM characterization of specimen deformed at 600 °C. (a) BF image revealing the DTs and SFs；(b) SAED pattern of the DTs；(c) the corresponding DF image of (a) indicating the DTs; (d) HRTEM of DTs and SFs; (e) BF of another location representing the abundant SFs; (f) SAED pattern of the SFs indicated by weak fringes; (g) high-density hierarchical SFs on two {111} slip planes (red arrows); (h) HRTEM image showing the hierarchical SF network (white dash line) and L-C locks (red dot) in different intersecting {111} slip systems.

Apart from the general dislocations, numerous SFs, nanotwins, as well as their extensive interactions of hierarchical SFs/DTs networks and LC locks, can cooperate with the deformation concurrently at RT. Such complex deformation substructure can be maintained at 600°C to enhance the strength and ductility during high temperature-tensile test. For traditional alloys, it is usually found that the twin strength is athermal, so higher temperatures will see dislocation plasticity. On the other hand, there has not been a consensus on twin strengthening in high entropy alloys. While some measurements suggest an athermal behavior, some other experimental evidences and modeling efforts show either moderate or strong temperature dependence [37]. Despite the debates, there has not been any evidence of elevated twin activities at elevated temperatures in high entropy alloys. Since twin nucleation stresses are usually at a few hundred MPas, the high strength here rather indicates that the high dislocation and SF densities lead to a higher probability of forming twin nuclei, so DTs are entangled with SFs and form complex networks here. Specifically, as verified by the TEM results for the high-temperature deformed samples, DTs not only provide favorable pathways to facilitate the dislocation motion for plasticity but also generate extra interfaces, leading to a pronounced dynamic Hall-Petch effect [36, 38], as well as another source of hetero-deformation-induced strengthening. Both factors increase the work-hardening rate, delay the onset of necking, and therefore increase the tensile ductility.

The TEM results also revealed a high density of SFs on multiple {111} planes in the samples deformed at both RT and 600 °C. Even though the strengthening effects of SFs may not be as notable as DTs, the massive SFs themselves can still contribute to the work hardening. For low SFE materials with larger separations between the dissociated partials, SFs afford an effective hardening mechanism for hindering dislocation motion and preventing recovery mechanisms such as cross-slip at elevated temperatures [39]. Additionally, the hierarchical SFs/DTs networks dynamically subdivide the matrix into massive nanodomains (with a length of 10~20 nm) during deformation.

Numerous immobile L-C locks along with nano-spaced SF networks are unexpectedly observed in NiCoCrAlTi alloy, which is rarely introduced in other fcc-type MEAs/HEAs under high-temperature tension. It is demonstrated that the generation of immobile L-C locks can act as notable obstacles to inhibit dislocation motion as well as can work as a Frank-Read source for dislocation multiplication, promoting steady and ascensive plastic deformation [40, 41]. Furthermore, L-C locks can generate and connect a high propensity of stable SFs as well as facilitates the establishment of SF networks, which can enhance their high structural stability to prohibit dissociations [42]. Moreover, the NiCoCrAlTi alloy can create extensive wide SFs, hence, should be more advantageous for generating the L-C locks. Consequently, L-C locks can accumulate dislocations and prohibit the motion of dislocations, which finally contributes to the outstanding strain-hardening capability in the NiCoCrAlTi alloy.

## 4. Summary

In this work, an excellent combination of mechanical properties with yield strength, UTS, and tensile ductility being ~1060 MPa, 1271 MPa, and 25%, respectively, was achieved in the NiCoCrAlTi HEA when deformed at an elevated temperature of 600 °C. The systematic characterization demonstrated that the planar SFs-based deformation substructure was mainly responsible for the enhanced strength-ductility synergy. Specifically, the DTs, dynamic high-density immobile L-C locks and hierarchical SFs/DTs networks resulting from extensive SFs intersections, and the ductile multi-component $L1_2$ precipitates cooperated simultaneously and substantially enhanced the mechanical properties of the NiCoCrAlTi alloy at high temperatures.

## CRediT authorship contribution statement

**Hongmin Zhang:** Investigation, Methodology, Writing – original draft, preparation. **Fanchao Meng**: Investigation, Writing – original draft. Haoyan Meng: Methodology. **Yang Tong:** Conceptualization. **Peter K. Liaw:** Writing - review & editing. **Xiao Yang:** Writing - review & editing. **Lei Zhao:** Methodology. **Haizhou Wang**: Writing - review & editing. **Yanfei Gao:** Conceptualization, Writing - review & editing. **Shuying Chen:** Conceptualization, Supervision, Writing – review & editing.

## Declaration of competing interest

The authors declare that they have no known competing financial interests or personal relationships that could have appeared to influence the work reported in this paper.

## Data availability

Data will be made available on request.

## Acknowledgments

This work was financially supported by the National Natural Science Foundation of China (No. 52001271). F. M. appreciates the support from the Natural Science Foundation of Shandong Province (ZR2021QE110). Y. T. acknowledges the financial support by Taishan Scholars Program of Shandong Province (tsqn202103052) and Yantai city matching fund for Taishan Scholars Program of Shandong Province. X. Y. was supported by Inner Mongolia Science and Technology Major Project (No. 2020ZD0011). Y. G. acknowledges fruitful discussions with W.A. Curtin.

101019.